\documentclass[times,twocolumn,final]{elsarticle}

\usepackage{cnf}

\usepackage{framed,multirow}

\usepackage{amssymb}
\usepackage{latexsym}
\usepackage{amsmath}
\usepackage{makecell}
\usepackage{multirow}
\usepackage{graphicx}
\usepackage{subfigure}
\usepackage{booktabs}
\usepackage{bm}
\usepackage{float}
\usepackage{tabularx}
\usepackage{pdfpages}
\usepackage{soul}
\usepackage{colortbl}
\usepackage{tikz}

\usepackage{url}
\usepackage{xcolor}
\definecolor{newcolor}{rgb}{.8,.349,.1}

\usepackage{hyperref}

\usepackage[switch,pagewise]{lineno} 

\journal{Combustion and Flame}

\begin{document}

\verso{Zhang et al.}

\begin{frontmatter}

\title{HiPrFlame-An \textit{ab initio} based real-fluid modeling approach for high-pressure combustion-I. Rationale, methodology, and application to laminar premixed flames}%

\author[1]{Ting \snm{Zhang}}
\author[1]{Tianzhou \snm{Jiang}}
\author[1]{Mingrui \snm{Wang}}
\author[1]{Hongjie \snm{Zhang}}
\author[1]{Ruoyue \snm{Tang}}
\author[1]{Xinrui \snm{Ren}}
\author[1,2]{Song \snm{Cheng}\corref{cor1}}
\cortext[cor1]{Corresponding author}
\emailauthor{songcheng@polyu.edu.hk}{Cheng Song}

\address[1]{Department of Mechanical Engineering, The Hong Kong Polytechnic University, Kowloon, Hong Kong SAR, PR China}
\address[2]{Research Institute for Smart Energy, The Hong Kong Polytechnic University, Kowloon, Hong Kong SAR, PR China}

\begin{abstract}
High-pressure combustion is central to modern propulsion and power-generation systems, where operating pressures often exceed the critical point of working fluids, resulting in pronounced real-fluid effects that fundamentally alter thermodynamic and transport properties. Existing methods for quantifying real-fluid behaviors typically rely on empirical correlations, fitted potentials, and(or) cubic equations of state (EoS), which lack the accuracy required for species coverage and extreme conditions encountered in combustion processes. As such, this study introduces HiPrFlame, a novel \textit{ab initio}-based modeling framework for high-pressure combustion, designed to deliver unprecedented fidelity in real-fluid property prediction in high-pressure combustion modeling. HiPrFlame integrates third-order Virial EoS derived from \textit{ab initio} intermolecular potentials, thereby real-fluid departure functions for real-fluid thermodynamics and Enskog theory for real-fluid transport properties, with all implemented within a versatile OpenFOAM architecture that can be used for 0-D to 3-D real-fluid modeling. To accelerate multidimensional simulations, artificial neural network surrogate models are trained on a comprehensive property database, enabling efficient real-fluid property updating. The framework is demonstrated through case studies of high-pressure hydrogen combustion, including homogeneous autoignition and one-dimensional laminar premixed flames. Results demonstrate that HiPrFlame accurately captures experimental data for both thermodynamic and transport properties, significantly outperforming traditional methods. Real-fluid effects are shown to enhance combustion reactivity, increasing, particularly, laminar flame speeds by over 400 \% even at subcritical pressures, which is much higher than previously observed in homogeneous reactors, highlighting the significant contributions from real-fluid transport. Comparative analyses reveal that accurate real-fluid modeling requires at least third-order Virial EoS based on \textit{ab initio} potentials, and shall not be applied based on lower-order or L-J potential-based approaches. HiPrFlame establishes a new standard for predictive high-pressure combustion modeling, providing critical insights into the role of real-fluid effects and offering a robust platform for future research. 
\end{abstract}


\end{frontmatter}


\section*{Novelty and significance statement}

This study presents HiPrFlame, a novel \textit{ab initio}-based modeling framework for high-pressure combustion that enables unprecedented accuracy in characterizing real-fluid effects. This is the first time that Enskog theory is coupled with high-order Virial EoS and \textit{ab initio} intermolecular potentials. By doing so, HiPrFlame surpasses existing approaches, as well as empirical EoSs and potentials in both real-fluid thermodynamic and transport property predictions. Moreover, HiPrFlame is implemented in OpenFOAM and enhanced with ANN surrogate models, offering versatile, high-fidelity real-fluid simulations from 0-D to 3-D. Demonstrated through high-pressure hydrogen combustion case studies, the framework reveals that real-fluid effects substantially promote reactivity, which can increase laminar flame speeds by over 400 \% even at subcritical pressures. This work is also the first study that quantifies and emphasizes, in high-pressure flame modeling, the necessity of high-order Virial EoS and \textit{ab initio} potentials for accurate real-fluid property computation.

\section{Introduction}
\label{intro}

High-pressure combustion brings significant enhancement in energy density and combustion efficiency to modern propulsion and power-generation systems. As such, the past decades have witnessed significant rise in pressure in these systems.  For instance, modern jet engines typically operate with maximum pressure over 4 MPa \cite{Epstein2012}, while more extreme conditions can be experienced in rocket engines (e.g., chamber pressure of the latest Raptor engine can reach 30 MPa \cite{Shriram2024}). These pressures could easily exceed the critical pressure of the working fluids (e.g., critical pressure of air is approximately 3.77 MPa), entering their trans-critical or super-critical regimes where the fluids exhibit considerably shifted thermodynamic and transport properties from those observed at low-pressure conditions. For instance, fluid density \cite{Sun2025} and surface tension \cite{Muller2016} can drop rapidly when reaching critical point. These shifted properties greatly influence combustion and emission characteristics, making high-pressure combustion fundamentally different and more complicated to understand than low-pressure combustion. 

Therefore, considerable efforts have been devoted to determine real-fluid properties in the past. As early as 1880, Van der Waals presented the corresponding state principle \cite{xiang2005}, which assumes that different substances in the same correspond pressure and temperature share the same corresponding state, where their thermophysical properties are corresponded to each other. The most important derivation of the corresponding state theory is the real-fluid departure function. Real-fluid departure function decouples real-fluid thermodynamic properties into two parts: (i) thermodynamic properties at low pressures; and (ii) their corrections at high pressures that are typically determined via a real-fluid Equation of State (EoS). Since the NIST-JANAF database \cite{Risch2021} has provided accurate thermodynamic properties at low pressures, the accuracy of the determined real-fluid thermodynamic properties is premised on the accuracy of the EoS used for real-fluid correction. In such regard,  considerable progress has been made in developing accurate real-fluid EoS. Today, many real-fluid EoS have been proposed with most being cubic EoS, including the Benedict-Webb-Rubin (BWR) EoS \cite{Younglove1987}, the Soave-Redlich-Kwong (SRK) EoS \cite{soave1972equilibrium} and the Peng-Robinson (PR) EoS \cite{Zips2018}. However, these EoS are semi-empirical and are mostly fitted on limited high-pressure experimental data, making them insufficient for applications under wider pressure and composition conditions, e.g., high-pressure combustion. The insufficiencies associated with the empirical, cubic EoS can be greatly mitigated by the Virial EoS \cite{hirschfelder1964molecular} which was derived based on the real-fluid partition function theory in statistical mechanics where real-fluid behavior is quantified via Virial coefficients. These Virial coefficients physically represent the intermolecular interactions in the fluid, which are computed from pre-determined intermolecular potentials. Computing the intermolecular potentials for mixtures involved in a combustion process can be quite challenging, making the application of Virial EoS in high-pressure combustion particularly challenging.

On the other hand, the determination of real-fluid transport properties (including viscosity, thermal conductivity, and mass diffusivity) has been more challenging, which has been heavily relying on experimental data. Generally, the transport properties can be further categorized: (i) viscosity and thermal conductivity which are caused by microscopic momentum transfer; (ii) mass diffusivity which are led by microscopic particle migration.

For computing real-fluid viscosity and thermal conductivity, various theories and empirical frameworks have been proposed \cite{Poling2001}. Among these frameworks, the two most popular methods have, perhaps, been the Chung method \cite{Chung1988} and the Locus method \cite{stephan2013viscosity}. Both methods can be applied to compute viscosity and thermal conductivity using real-fluid parameters like critical temperature and pressure. However, these methods are quite complicated to use and apply for a wide range of species. The Locus method \cite{stephan2013viscosity}, for instance, replies on solving more than 10 equations with 14 empirical constants. The Chung method \cite{Chung1988} presents a relatively simpler equation form but has 30 empirical constants. For computing mass diffusivity, the Takahashi method \cite{Takahashi} and the Riazi-Whitson method \cite{riazi1993estimating} are most popular. Same as the Chung and Locus methods, both Takahashi and Riazi-Whitson methods require determining numerous equations and constants.
Furthermore, these constants are empirically determined and might lead to significant discrepancies when handling complex molecular structures at extreme conditions, as typically seen in high-pressure combustion. In comparison, the Enskog theory \cite{Jervell2023} presents a non-empirical method directly derived from the Boltzmann equation and is capable of computing all three transport properties. In Enskog theory, real-fluid transport properties are determined based on Virial coefficients. Previous studies \cite{Millat1996} have already demonstrated the advantages of the Enskog theory than empirical methods in computing real-fluid transport properties, particularly for small molecules that can be safely approximated as hard spheres. The order of Enskog theory, hence the accuracy of the computed real-fluid transport properties, is directly correlated to the order of the Virial coefficients. Nevertheless, there have been no studies that combine Enskog theory with high-order Virial EoS (e.g., > $\rm{2^{nd}}$ order) for understanding real-fluid combustion.

Laminar flame speed (LFS) is an intrinsic property that is crucial to understanding high-pressure combustion. It is critical for determining flame kernel initiation and has been widely adopted as a standard target for developing and validating chemistry models at high temperatures. As such, there have been numerous efforts in measuring the LFS at high pressure conditions. Tse et al. \cite{Tse2000} investigated flame propagation at pressures up to 6 MPa using two concentric cylindrical vessels. Results indicated that flame instabilities dominated flame dynamics at these elevated pressures, especially the development of hydrodynamic cells. Bradley et al. \cite{Bradley2007} measured the high-pressure LFS of lean premixed hydrogen-air mixtures at equivalence ratios between 0.3 and 1.0. Due to the experimental difficulties, the highest pressure achieved was around 1 MPa. Despite the importance, measurements of LFS at high pressure  conditions are scarce and remain unavailable for most fuels \cite{Hayakawa2015, konnov2018}.

The limited experimental data for high-pressure LFS has also limited the development of accurate numerical models for quantifying real-fluid effects. Liang et al. \cite{Liang2019} numerically studied the influences of real-fluid properties on LFS of hydrogen/air mixtures. Real-fluid thermodynamic properties were determined using departure functions with SRK EoS, while real-fluid transport properties were determined via the Ely method \cite{Ely1981} and Takahashi method \cite{Takahashi}. They eventually found that the real-fluid thermodynamic properties impose the greatest impact on the simulated LFS due to the corresponding reduction of adiabatic flame temperature, while the real-fluid transport properties exhibit relatively smaller impacts. Lv et al. \cite{Lv2025} investigated the real-fluid effects on modeling laminar premixed $\rm{H_2/O_2}$ flames under cryogenic and high-pressure conditions. In their study, real-fluid thermodynamic properties were also computed using departure functions with RK EoS, while the real-fluid transport properties were determined using the Chung method \cite{Chung1988} and the Takahashi method \cite{Takahashi}. Although the calculated specific heat capacity of hydrogen deviated considerably from the NIST data, their results revealed that the correction of real-fluid thermodynamics plays a critical role in the prediction of flame structure and the LFS, while the correction of transport properties is critical for predicting flame thickness.
Zhang et al. \cite{Zhang2024a} numerically investigated the propagation of laminar oxy-syngas and oxy-methane flames diluted by supercritical carbon dioxide and incorporated the same real-fluid models as Lv et al \cite{Lv2025}. They found that the relative uncertainty on the simulated LFS caused by real gas effects and by different chemistry models can be on the same order of magnitude. Terashima et al. \cite{Terashima2023} established a real-fluid computational fluid dynamics modeling for high-pressure laminar premixed $\rm{H_2/O_2}$ propagating flames where real-fluid effects were described via the departure functions and SRK EoS for thermodynamics and the Chung method \cite{Chung1988} and Riazi-Whitson method \cite{riazi1993estimating} for transport properties. They subsequently conducted a quantitative analysis of real-fluid effects on chemical kinetics via modified equilibrium constant for chemical kinetics. The results indicated the significance of real-fluid effects, which highly depend on the species composition and thermodynamic conditions.

Studies mentioned above consistently highlighted the significant impact of real-fluid behaviors on modeling high-pressure flames, highlighting the need for adequately representing real-fluid effects. Nevertheless, the real-fluid modeling frameworks adopted in existing studies, as summarized in Table \ref{tab1}, remain empirical, with inconsistent results reported. As a result, simulations of high-pressure laminar flame measurements in previous studies were mostly conducted by assuming ideal gas behavior, with the real-fluid effects completely overlooked. This has introduced significant errors into the reported simulation results, which will also be propagated to the developed chemistry and transport models that have been widely used for predictive combustion modeling.

\begin{table*}[!t]
\caption{\label{tab1}Real-fluid models adopted in existing studies for modeling high-pressure laminar premixed flames.}
\centering
\fontsize{8pt}{9pt}\selectfont
\begin{tabular}{|c|c|c|c|}
\hline
Ref & Thermodynamics & Transport (viscosity and thermal conductivity) & Transport (mass diffusivity)\\
\hline
Liang et al. \cite{Liang2019} & SRK	& Ely &	Takahashi  \\
\hline
Lv et al. \cite{Lv2025} & PR &	Chung	& Takahashi \\
\hline
Zhang et al. \cite{Zhang2024a} & RK &	Chung &	Takahashi \\
\hline
Terashima et al. \cite{Terashima2023} & SRK	& Chung &	Riazi-Whitson \\
\hline
\end{tabular}
\end{table*}

Aware of these inadequacies, Cheng and co-workers pioneered the establishment of a robust real-fluid modeling framework \cite{Wang2025} that couples non-empirical high-order Virial EoS (up to $\rm{8^{th}}$ order), \textit{ab initio} multi-body intermolecular potential, and real-fluid governing equations. They found that the developed framework impressively replicates the real-fluid thermodynamic properties (predicting identical results with the experimental measurements) and that Virial EoS shall not be determined based on empirical intermolecular potentials such as the Lennard-Jones (L-J) potential. The framework was further applied to simulate real-fluid autoignition in rapid compression machines \cite{Wang2025d} and shock tubes \cite{Wang2025, Wang2025e}, as well as real-fluid pyrolysis and oxidation in jet-stirred reactors \cite{Wang2025a, Wang2025b} and flow reactors \cite{Wang2025f} at high-pressure conditions, where significant real-fluid effects were observed in all reactors (the errors introduced by ignoring real-fluid behaviors at high-pressure conditions are substantially higher than typical levels of measurement uncertainties). As a continuation of this dedicated effort, this study aims to extend the developed real-fluid modeling framework from 0-D simulations to 1-D to 3-D simulations by establishing a new architecture in OpenFOAM, named HiPrFlame, and demonstrate its capability via a case study of high-pressure hydrogen laminar premixed flames. Remarkably, the HiPrFlame will also enable 0-D simulations for high-pressure pyrolysis, oxidation and autoignition with consistent solver setting and computational speed-up.

\section{Methodology and rationale}

\begin{figure*}[!t]
\centering
\includegraphics[scale=0.8]{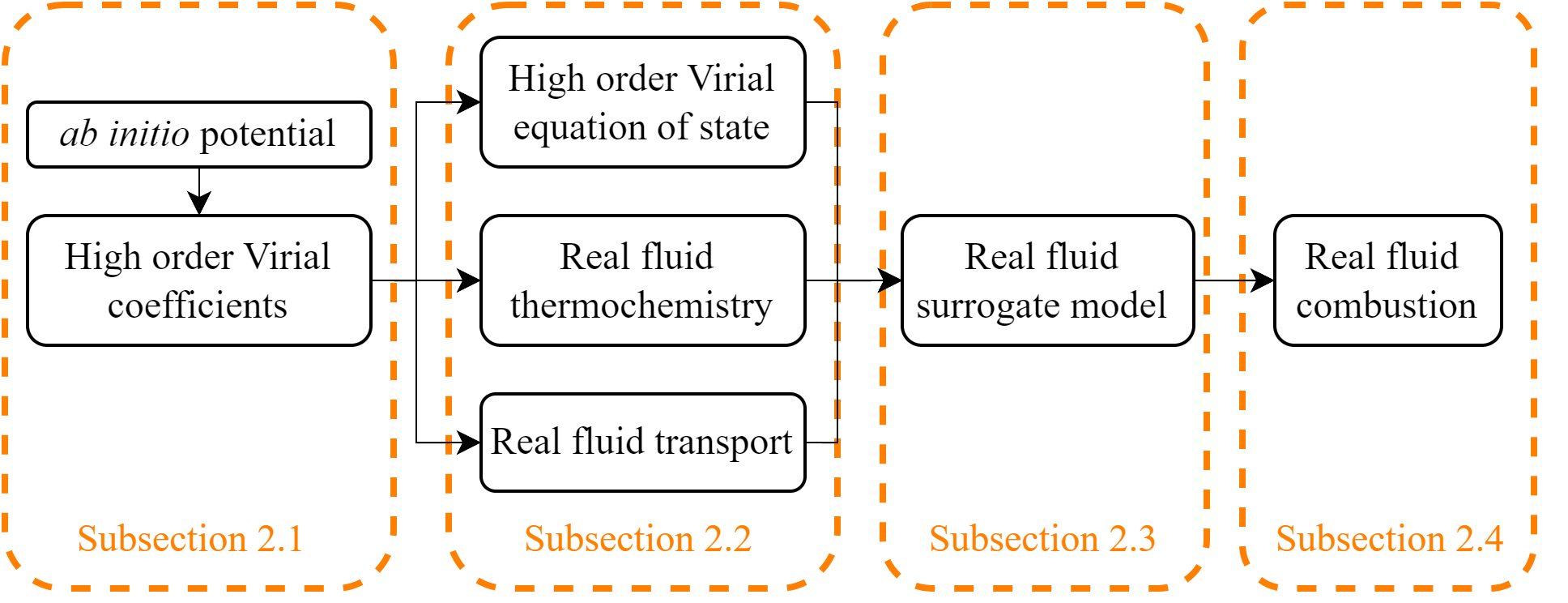}
\caption{The structural overview of HiPrFlame.} \label{research_plan}
\end{figure*}

Fig. \ref{research_plan} shows a structural overview of HiPrFlame, encompassing four main stages: (i) High-order mixture Virial EoS determination, (ii) real-fluid properties determination; (iii) surrogate models for real-fluid properties; and (iv) real-fluid combustion modeling, which are discussed in Sections \ref{section-2-1} to \ref{section-2-4}, respectively.

\subsection{High-order mixture Virial EoS determination} \label{section-2-1}
The Virial EoS can be used with different orders, with higher order typically demonstrating better accuracy, since the $\rm{N_{th}}$ order Virial coefficient physically represents the contribution of intermolecular interactions between $\rm{N}$ molecules in the mixture to system properties. In our previous studies, we have demonstrated that the Virial EoS should be used at the $\rm{3^{rd}}$ order in order to ensure its accuracy and advantage to cubic EoSs \cite{Wang2025} while the influences of intermolecular interactions begin to saturate at the 4th order and above. Therefore, $\rm{3^{rd}}$ order mixture Virial EoS will be adopted in this study. The $\rm{3^{rd}}$ order Virial EoS can be written as:

\begin{equation} \label{eq1}
    \frac{p}{\rho R T}=1+B(T)\rho +C(T) \rho^2 
\end{equation}
where $R$ is the specific gas constant, and $B$ and $C$ are second and third virial coefficients in the unit of $\rm{m^3/kg}$ and $\rm{(m^3/kg)^2}$, respectively. As can be seen from Eq. \ref{eq1}, if no intermolecular interactions are considered, the $\rm{2^{nd}}$ and $\rm{3^{rd}}$ order Virial coefficients become zero, and the Virial EoS reduces to the ideal EoS.

Virial coefficients can be computed from intermolecular potentials. To date, high-order Virial coefficients have already been computed based on \textit{ab initio} intermolecular potentials for species such as $\rm{H_2}$ \cite{Patkowski2008,Garberoglio2013}, $\rm{N_2}$ \cite{Hellmann2013}, $\rm{O_2}$ \cite{Hellmann2023} and $\rm{H_2O}$ \cite{Garberoglio2018a}. In this study, the existing data of Virial coefficients are fitted into polynomials as functions of temperature. To further demonstrate whether this is the case for transport properties, L-J potentials are also adopted in this study for quantitative comparison with \textit{ab initio} potentials (discussed later).

\begin{figure*}[h]
\centering
\includegraphics[width=0.9\textwidth]{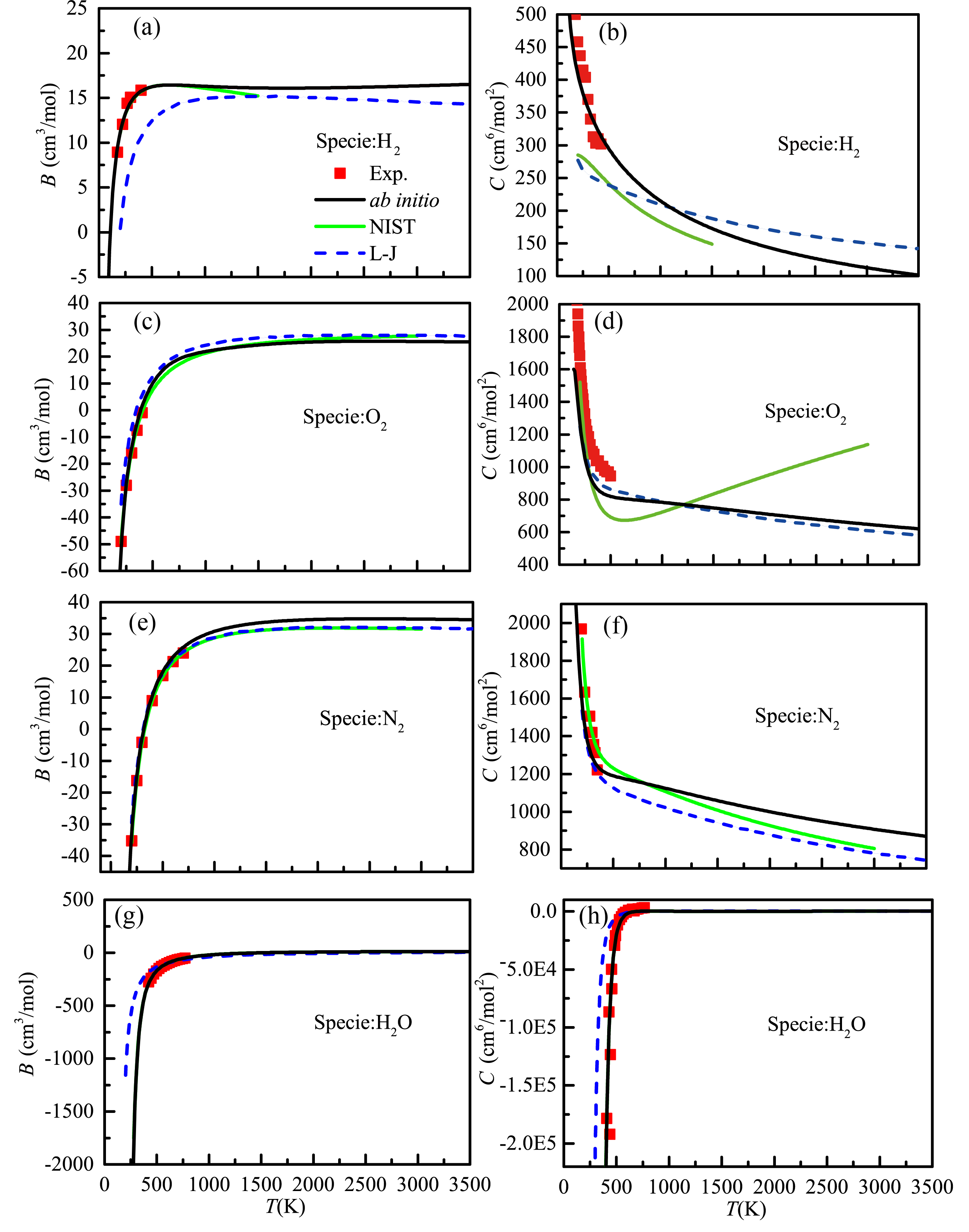}
\caption{The Virial coefficients of $\rm{H_2}$ (a-b), $\rm{O_2}$ (c-d), $\rm{N_2}$ (e-f), and $\rm{H_2O}$ (g-h) using different intermolecular potentials, along with the measurements from \cite{Goodwin1964, Kusalik1995, Sevastyanov}.}\label{virial-all}
\end{figure*}

To determine the Virial coefficients, the mixture species (e.g., a chemistry model) need to be pre-defined. In this study, the chemistry model proposed by Conaire et al. \cite{OConaire2004} is selected, which includes 10 species. The $\rm{2^{nd}}$ and $\rm{3^{rd}}$ order Virial coefficients computed using different potentials are illustrated in Fig. \ref{virial-all}, along with the experimental data reported in \cite{Goodwin1964, Kusalik1995, Sevastyanov} and those from the NIST database \cite{Huber2022}. It is evident from Fig. \ref{virial-all} that the Virial coefficients computed from the \textit{ab initio} potentials perform the best in capturing the experimental data, while the L-J potential performed the worst. Surprisingly, there are noticeable discrepancies between the experiments and the NIST database. This is particularly obvious for the $\rm{3^{rd}}$ order Virial coefficients, highlighting the necessity to determine these Virial coefficients based on \textit{ab initio} intermolecular potentials. 

The details about the fitting and determination of Virial coefficients can be found in the Supplementary Material.

\subsection{Real-fluid properties determination} \label{section-2-2}
\subsubsection{Thermodynamic properties}
With the high-order Virial EoS from Section \ref{section-2-1}, the real-fluid thermodynamic properties can be determined using the departure functions. For instance, the departure function for enthalpy, derived based on the $\rm{3^{rd}}$ order Virial EoS can be expressed as:

\begin{equation}
\begin{aligned}
\frac{\Delta h}{RT}=&(\frac{p}{RT})(B-T\frac{dB}{dT}) \\
+&(\frac{p}{RT})^2(C-\frac{T}{2}\frac{dC}{dT}-B^2+B\frac{dB}{dT})
\end{aligned} 
\end{equation}

\begin{figure}[h]
\centering
\includegraphics[scale=0.65]{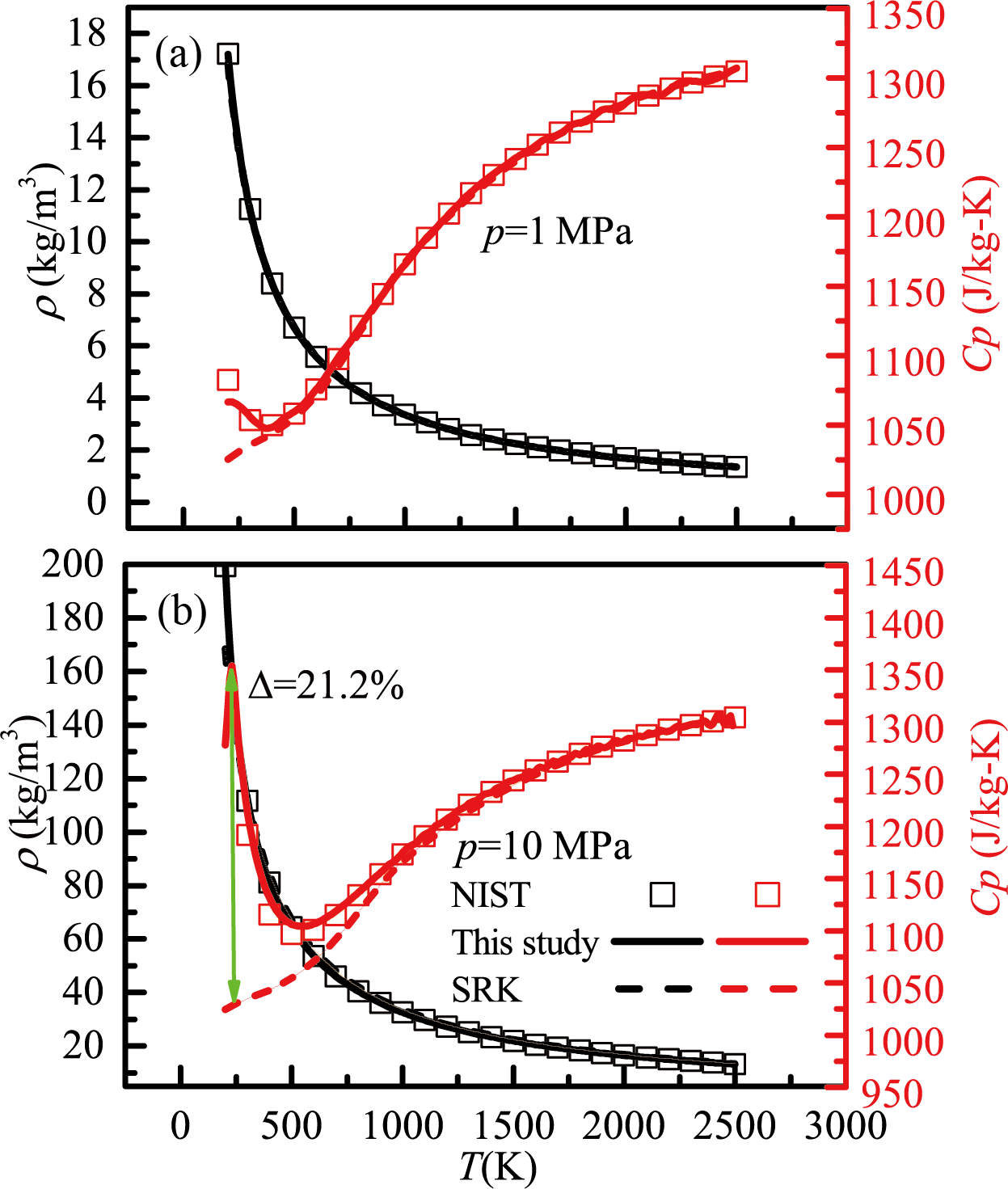}
\caption{Density and heat capacity of N2 at different pressures and temperatures, computed using the $\rm{3^{rd}}$ order Virial EoS (solid lines) and the SRK cubic EoS (dashed lines), along with the experimental data from NIST \cite{Huber2022} (a) 1 MPa; (b) 10 MPa.} \label{N2-thermo}
\end{figure}

With the ideal gas properties, the real fluid enthalpy can be determined as

\begin{equation}
   h=h_{ig}+\Delta h 
\end{equation}
where $h_{ig}$ is the ideal gas enthalpy. Similarly, all other real-fluid thermodynamic properties can be computed following this approach, for which the details can be found in our previous work \cite{Wang2025}. Fig. \ref{N2-thermo} summarizes the computed density and heat capacity of $\rm{N_2}$ from using the $\rm{3^{rd}}$ order Virial EoS and the SRK EoS at different pressures as functions of temperature, along with the experimental data from NIST database \cite{Huber2022}. It can be seen from Fig. \ref{N2-thermo} that, although both EoSs predict similar density that agree well with experimental data, the discrepancy in predicted heat capacity is obvious. Specifically, the $\rm{3^{rd}}$ order Virial EoS excellently captures the experimental data for heat capacity, while the SRK EoS underpredicts the experiments within the low-temperature regime with greater disagreements observed at the higher-pressure condition (up to 21.2 \%).

\subsubsection{Transport properties}

In this study, the real-fluid transport properties will be computed using the Enskog theory in conjunction with $\rm{3^{rd}}$ order Virial EoS, which has not been achieved in the past. According to Enskog theory, the real-fluid dynamic viscosity ($\mu$), thermal conductivity ($\lambda$), and mass diffusion ($D$) can be determined as,

\begin{equation}
\frac{\mu}{\mu_{ig} \rho b}=\frac{1}{y}+0.8+0.761y
\end{equation}

\begin{equation}
\frac{\lambda}{\lambda_{ig} \rho b}=\frac{1}{y}+1.2+0.755y
\end{equation}

\begin{equation}
\frac{pD}{(pD)_{ig} \rho b}=\frac{1}{y}+1
\end{equation}
where the subscript $ig$ denotes ideal-gas transport properties at the atmospheric pressure, and $b$ and $y$ are correction factors that can be computed as,

\begin{equation}
   b=B+T\frac{dB}{dT}
\end{equation}

\begin{equation}
y=\frac{1}{\rho RT}\left[T\left(\frac{\partial p}{\partial T}\right)_{\rho}\right]-1
\end{equation}
where $\frac{\partial p}{\partial T}$ should be computed based on the $\rm{3^{rd}}$ order Virial EoS, which has been detailed in our previous study \cite{Wang2025}.

\begin{figure}[H]
\centering
\includegraphics[scale=0.65]{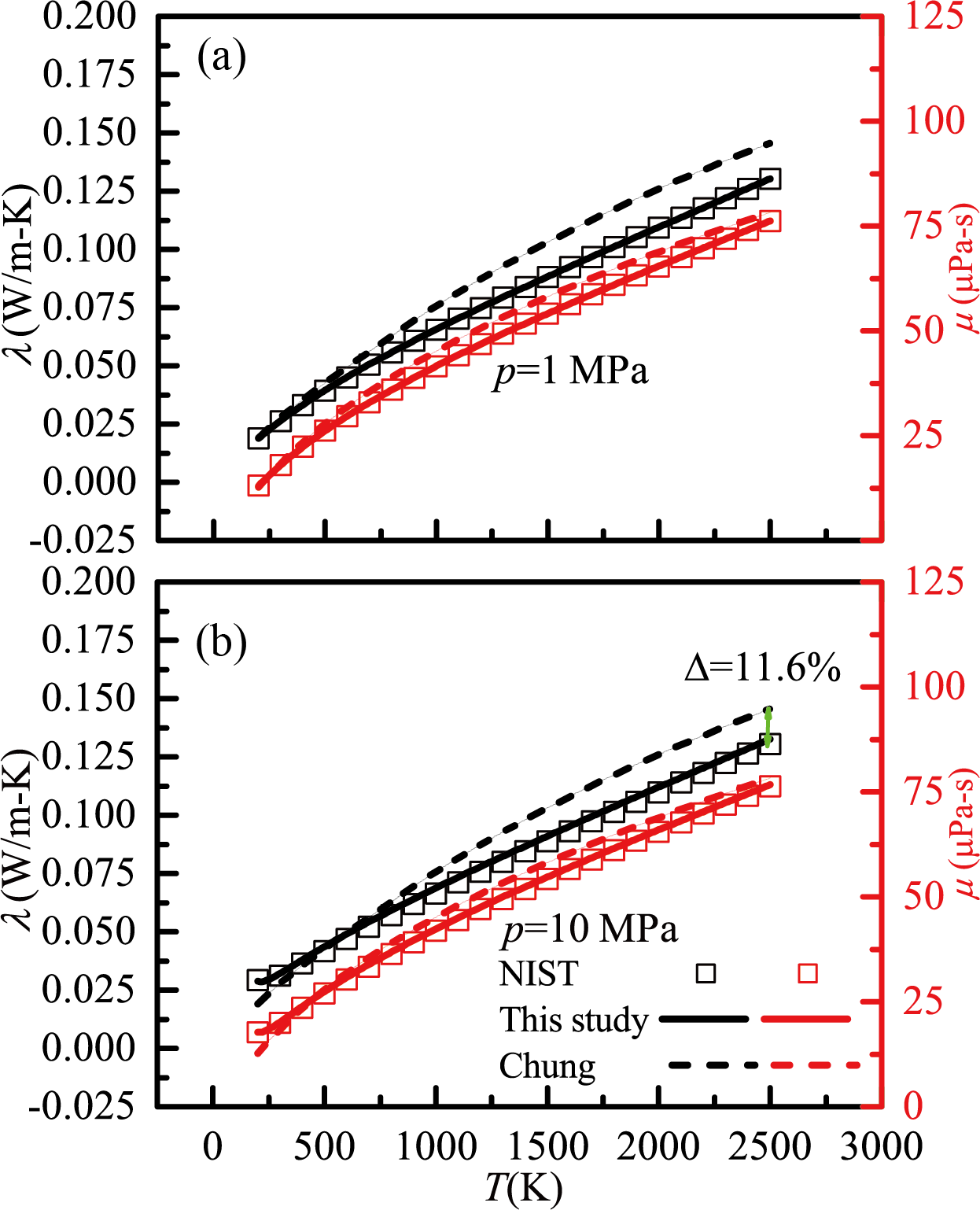}
\caption{Thermal conductivity and viscosity of $\rm{N_2}$ at different pressures and temperatures, computed using the Enskog theory coupled with $\rm{3^{rd}}$ order Virial EoS (solid lines) and the Chung method (dashed lines), along with the experimental data from NIST \cite{Huber2022} (a) 1 MPa; (b) 10 MPa.} \label{N2-trans}
\end{figure}

The computed real-fluid conductivity and dynamic viscosity using the method proposed in this study and the Chung method \cite{Chung1988} for $\rm{N_2}$ are shown in Fig. \ref{N2-trans}. As shown in Fig. \ref{N2-trans}, the transport properties predicted using the novel approach developed in this study agree impressively with the experiments, while those predicted using the Chung method consistently overpredict the experiments. Over the whole temperature and pressure range studied, the highest prediction error of the proposed method is within 1.6 \%, while that of the Chung method can reach 11.6 \%.

\subsubsection{Real-fluid mixing rules}
Mixing rules need to be determined to compute mixture real-fluid properties from pure substance real-fluid properties. This is crucial as it has been found that simple mixing rules based on species mass fraction weighted average do not work in high-pressure combustion \cite{Nguyen2022}. In this study, the mixing rules proposed and validated in our previous work \cite{Wang2025} are implemented.

\subsection{Surrogate models for real-fluid properties}\label{section-2-3}
Solving real-fluid properties in real-fluid combustion modeling can be approached either analytically or using the look-up table method. The former, which was adopted in our previous 0-D modeling framework (i.e., the UHPC-RF-Master package) \cite{Wang2025} can be computationally expensive to implement in high-dimensional modeling (e.g., expensive derivatives for the mixture Virial coefficients, which are composition- and temperature-dependent), while the latter uses a well tabulated high-dimensional property dataset and can considerably reduce the computational cost \cite{Zhanga,Liu2014a}. In this study, the look-up table method is adopted, with machine learning frameworks further used to construct the surrogate models for the high-dimensional property dataset. 

As discussed in Section \ref{section-2-2}, ideal gas properties are needed to compute the real-fluid properties. To this end, an ideal gas property dataset is first constructed via sampling. Since the concentration of species varies within a specific range during combustion, Latin Hypercube Sampling (LHS) \cite{Helton2003} is used to randomly generate mixture samples within a prescribed concentration range for each of the 9 species in the chemistry model. Then the species concentrations are normalized to ensure the sum of the concentrations of all species equals to 1. During sampling, temperature is varied by 5 K from 200 K to 3500 K. For the laminar premixed flame simulations in this study, pressure can be assumed constant \cite{Designs2023} and only real-fluid properties at different temperature need to be cmputed. However, for autoignition simulations in this study, variation in pressure needs to be considered. As such, the pressure is varied by 0.1 MPa from 0.1 MPa to 10 MPa. At each temperature and pressure condition, mixture composition is sampled, and the corresponding ideal gas properties are determined. Thereafter, the real-fluid properties are further determined using the methods discussed in Section \ref{section-2-2}. To eliminate singularities in the dataset (e.g., heat capacity becomes infinitely large near critical point and saturation point) and avoid biased training (e.g., the range of values vary significantly for different properties), 3-$\sigma$ principle \cite{Zhou2021} and Z-score normalization \cite{Henderi2021} are applied, respectively, to preprocess the property dataset before training. The final property dataset contains more than 1,300,000 data points. 

\begin{figure}[h]
\centering
\includegraphics[scale=0.5]{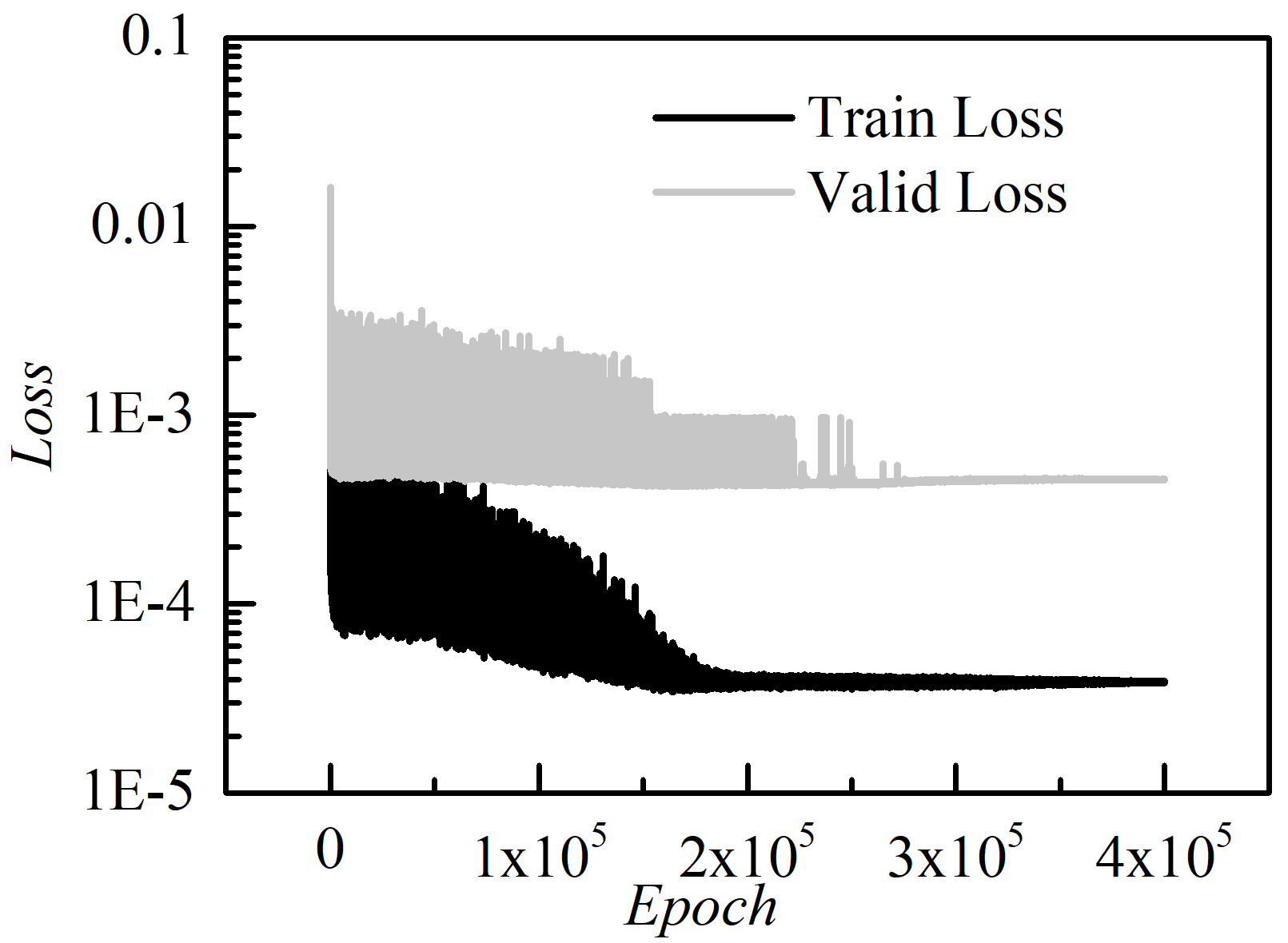}
\caption{Loss curves per epoch for the test and train set of the ANN model.} \label{loss}
\end{figure}

The Artificial Neural Networks (ANN) model used in this study contains 5 hidden layers and each hidden layer contains 256 nodes. 1000 random data points are selected as the test set. During training, Huber loss function \cite{huber1992robust} is adopted to assess training performance, with Adam optimization \cite{Kingma2017} used for adaptive learning and cosine annealing \cite{CosineAnnealingLR} adopted to adjust the learning rate dynamically. The training code is developed based on \textit{PyTorch} \cite{Paszke2019}. Fig. \ref{loss} demonstrates the training and testing loss as function of epochs. The train loss decreases from 0.01 to $2\times10^{-5}$ within $2\times10^5$ epochs and the validation loss decreases from 0.01 to $4\times10^{-4}$ within $3\times10^5$ epochs. Both loss curves converge within $3\times10^5$ epochs, indicating the robustness of the training process.

\begin{figure}[H]
\centering
\includegraphics[scale=0.65]{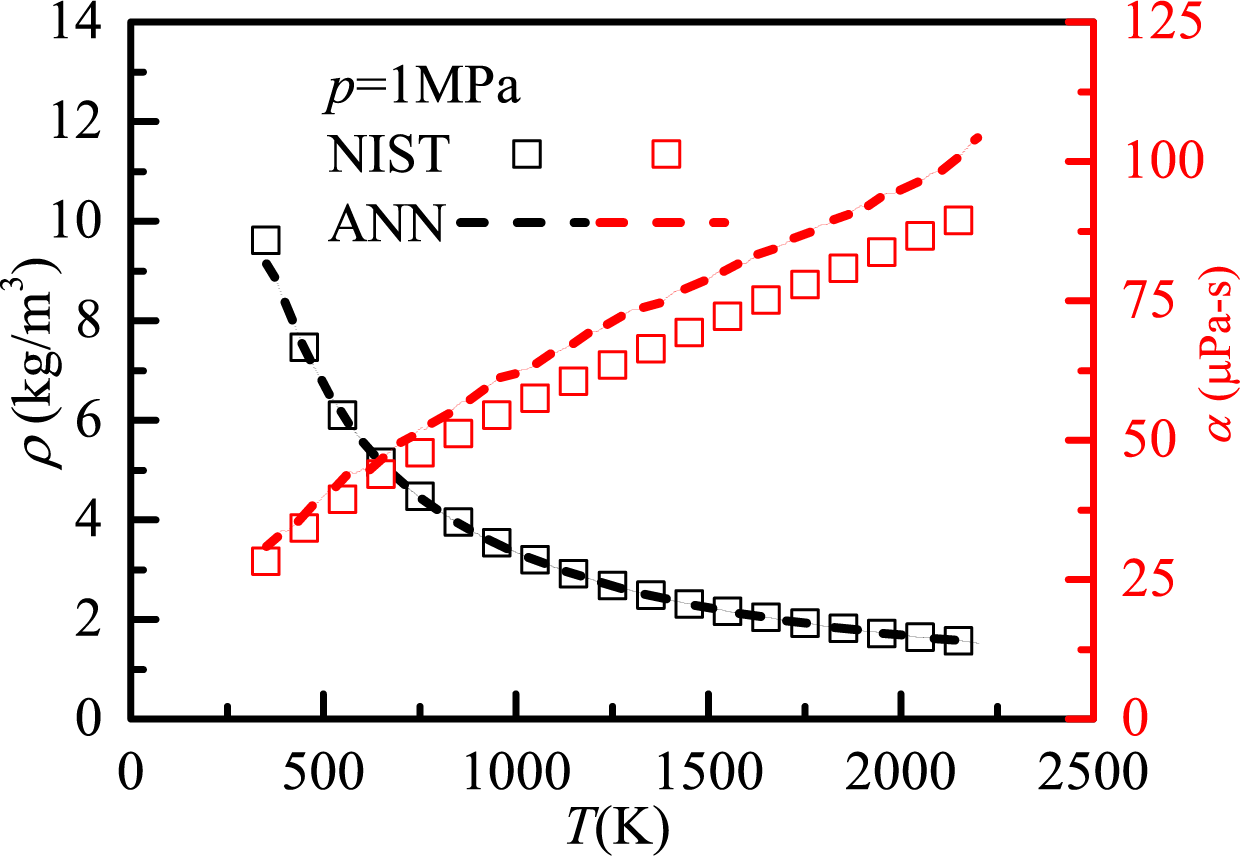}
\caption{Density and thermal diffusivity of $\rm{N_2}$ predicted by the trained ANN model at 1MPa, along with the experimental data from NIST \cite{Huber2022}.} \label{1Mpa-prop}
\end{figure}

Fig. \ref{1Mpa-prop} compares the density and thermal diffusivity of $\rm{N_2}$ computed from the trained ANN model against the NIST data. It can be seen from Fig. \ref{1Mpa-prop} that the trained ANN model achieves great agreement with experiments, reaching an error of less than 2 \% for density and less than 10 \% for thermal diffusivity at the conditions studied.

\subsection{Real-fluid combustion modeling} \label{section-2-4}
\subsubsection{Conservation equations}\label{section-2-4-1}

To solve transient compressible reacting flow with N species, the conservation equations of mass and momentum are given by:
\begin{equation}
  \frac{\partial \rho}{\partial t}+ \nabla \cdot (\rho \mathbf{U})=0 
\end{equation}

\begin{equation}
\frac{\partial \rho \mathbf{U}}{\partial t} + \nabla \cdot (\rho\mathbf{UU})=-\nabla p +\nabla\cdot\tau
\end{equation}
where $\mathbf{U}$ is the velocity victor, and $\tau$ is the viscous stress tensor. For compressible Newtonian flow, $\tau$ is given by:

\begin{equation}
\tau=\mu\left[\nabla\mathbf{U}+(\nabla\mathbf{U})^{\mathrm{T}}\right]-\frac{2}{3}\mu(\nabla\cdot\mathbf{U})\mathbf{I}
\end{equation}
where $\mu$ is the dynamic viscosity and $\mathbf{I}$ is the unit tensor.

The mass fraction of each species $Y_i$ is calculated through reserving the species transport equations,
\begin{equation} \label{eq17}
\frac{\partial \rho Y_i}{\partial t} + \nabla \cdot (\rho\mathbf{U} Y_i)= -\nabla \cdot (\rho \mathbf{V}_i) + \omega_i
\end{equation}
where $\omega_i$ is the net production rate of species, Vector $\mathbf{V}$ is the mass diffusion flux vector of species $i$ is given by Fick's Law and with a correction velocity to ensure mass conservation.

\begin{equation}
   \mathbf{V}_i=- D_{mix} \nabla Y_i + Y_i \sum_{i=1}^{N} (D_{mix} \nabla Y_i)
\end{equation}
where $D_{mix}$ is the mixture averaged diffusion coefficient. Eq. \ref{eq17} gives N-1 equations for N-1 species, with an additional equation from the following

\begin{equation}
    Y_N=1-\sum_{i=1}^{N-1}Y_i
\end{equation}

The conservation of energy in this work is established based on total specific enthalpy instead of sensible specific enthalpy:
\begin{equation}
   \frac{\partial \rho h}{\partial t}+\nabla \cdot (\rho \mathbf{U} h)+\nabla \cdot (\rho \mathbf{U} K) = \frac{\partial p}{\partial t} - \nabla \cdot \mathbf{q} 
\end{equation}
where $K$ denotes the local kinetic energy, $K=\frac{1}{2} \mathbf{U}^2$ , and $\mathbf{q}$ is the heat flux vector, which can be determined as:

\begin{equation}
  \mathbf{q} = -\frac{\lambda}{C_p} \nabla h + \rho \sum_{i=1}^{N}h_i Y_i \mathbf{V}_i
\end{equation}
where $\alpha$ is the specific thermal diffusivity (i.e., $\frac{\lambda}{C_p}$), where $\lambda$ is thermal conductivity and $C_p$ is specific thermal capacity.

\subsection{Implementation in OpenFOAM}

\begin{figure}[h]
\centering
\includegraphics[width=0.5\textwidth]{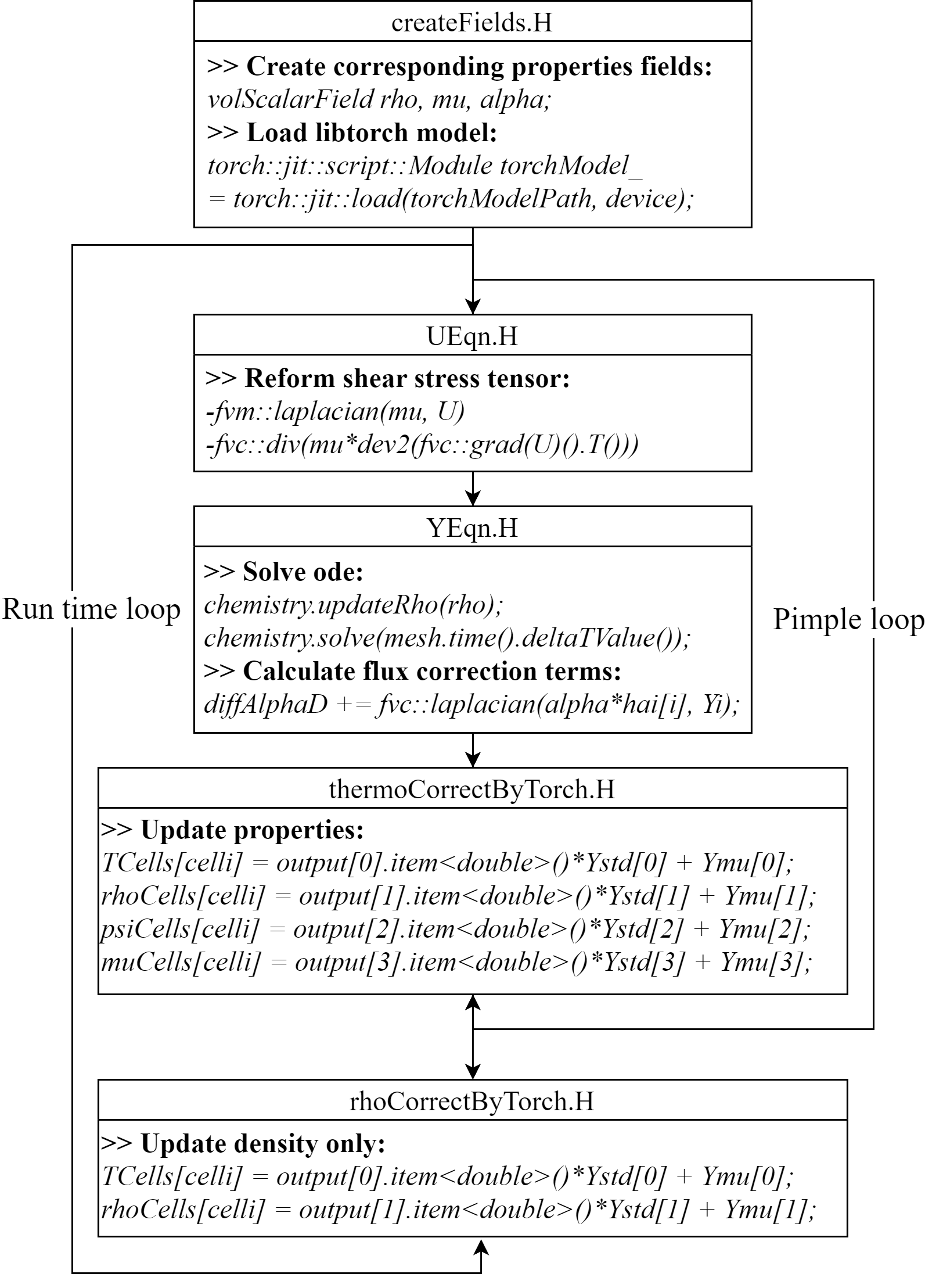}
\caption{Schematic of the front-end mechanism for real-fluid property updating in HiPrFlame.} \label{implimentation}
\end{figure}

The conservation equations in Section \ref{section-2-4-1} are implemented into open source CFD toolbox OpenFOAM v7.0 \cite{Greenshields2019}. Since OpenFOAM adopts a uniform mesh structure, one can easily convert between 0-D to 3-D by changing the boundary conditions. During each iteration of computation, all thermodynamic and transport properties in Eq. 14-21 are corrected using the real fluid properties predicted from the ANN model (see Section \ref{section-2-3}). In this work, a front-end mechanism is established to realize thermal physical property updating without accessing Runtime Type Selection (RTS) model \cite{Maric2021}. Fig. \ref{implimentation} illustrates the schematic of the front-end mechanism. Specifically, in \textit{createFields.H}, scalar fields for all properties, including density, viscosity, thermal diffusivity, mass diffusivity and enthalpy, are declared. Then the ANN model trained in Section \ref{section-2-3} is loaded using libtorch. In \textit{UEqn.H}, the compressible shear stress tensor is reformed to exclude turbulence terms. In \textit{YEqn.H}, the density field and the solver setting such as local time step are passed to the ODE solver to solve the source term $\omega_i$ in Eq. \ref{eq17}. At the end of every Pimple loop, \textit{thermoCorrectByTorch.H} is included to update all thermodynamic and transport properties. Similarly, at the end of every run time loop, \textit{rhoCorrectByTorch.H} is included to update the density field. The ODE solver used in this study is developed based on Deepflame \cite{Mao2023} interface between Cantera and OpenFOAM.

\subsubsection{Numerical setup for 1-D laminar premixed flames}
In this study, the second order upwind scheme is applied to the discretization of most convection terms, while for discretization of shear stress, the second order central difference scheme is employed. As for time discretization, the Euler scheme is applied, which is a first order implicit scheme. The Pressure Implicit with Splitting of Operators (PISO) method is used to deal with pressure-velocity coupling. Detailed description about the numerical schemes mentioned above can be found in \cite{Moukalled2016}.

\begin{figure}[H]
\centering
\includegraphics[scale=0.8]{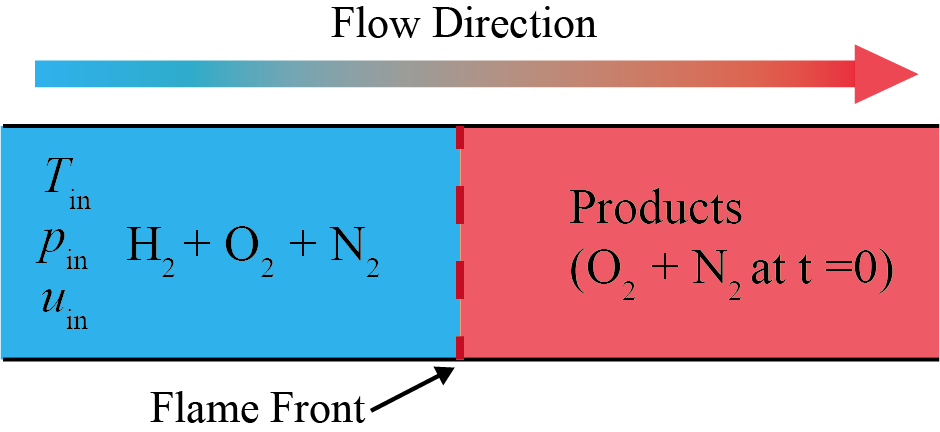}
\caption{Schematic of the computational domain for 1-D freely propagating laminar flames.} \label{initial_condition}
\end{figure}

To simulate 1-D freely propagating laminar flames, a 0.56 m long computational domain is set up, as shown in Fig. \ref{initial_condition}. 17,000 grids are generated, and mesh near the flame front (defined as the position of $\nabla T_{max}$) is refined to ensure simulation accuracy. Upstream of the flame front is defined as a homogeneous mixture with a specific air/fuel ratio, temperature and pressure, while downstream of the flame front is set as a mixture of oxygen and nitrogen with an ignition temperature higher than 2200 K. After obtaining transient simulation results, the LFS, $S_L$, is determined by subtracting the propagation velocity of flame front  by the inlet velocity. 

\section{Results and discussion}

\subsection{Comparison between potentials and Virial orders} \label{section-3-1}
\begin{figure*}[h]
\centering
\includegraphics[width=\textwidth]{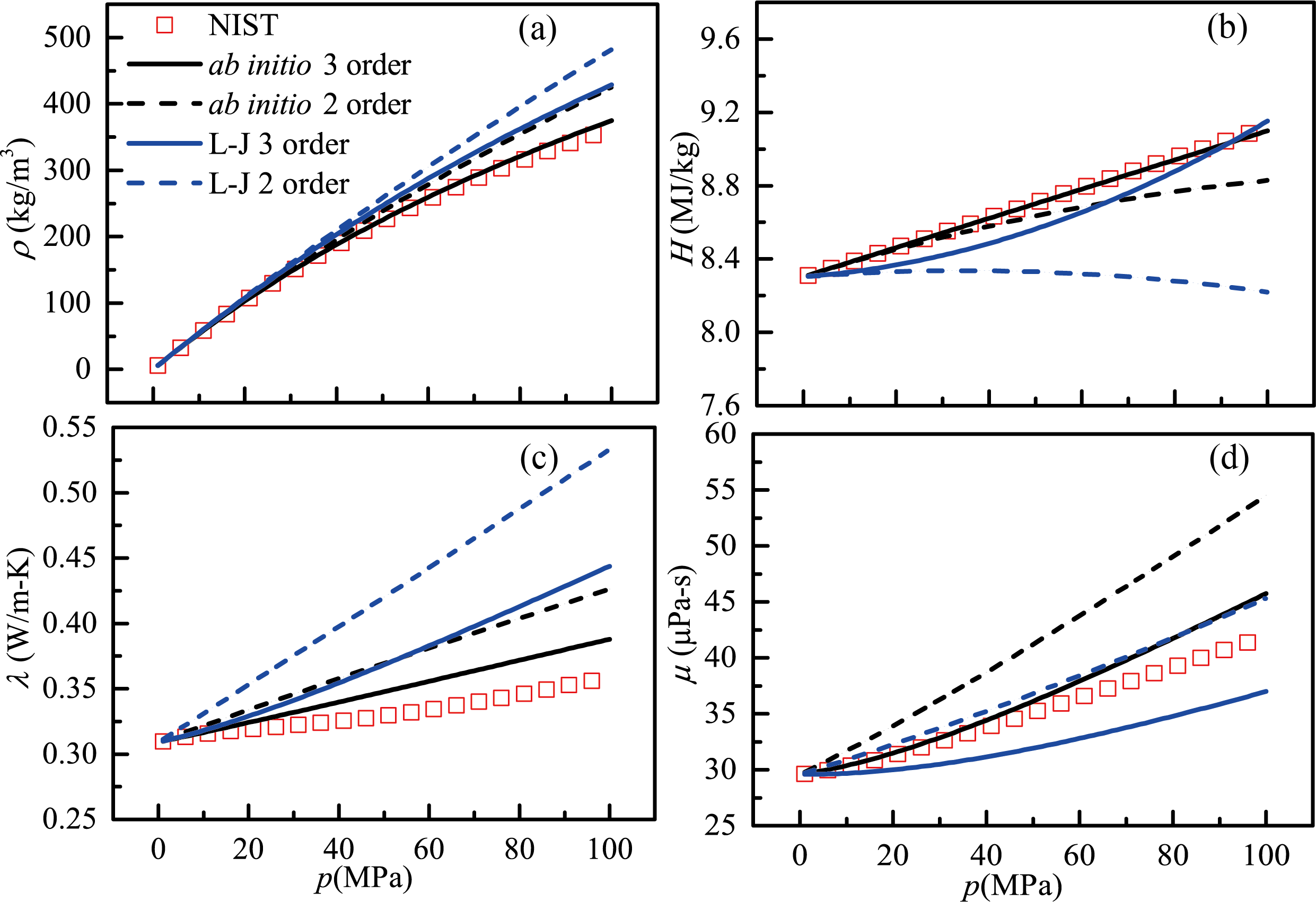}
\caption{Thermodynamic and transport properties of $\rm{N_2}$ at 600 K and 1-100 MPa computed using the Enskog theory coupled with Virial EoS based on different intermolecular potentials and Virial orders, along with the experimental data from NIST \cite{Huber2022} (a) density; (b) enthalpy; (c) thermal conductivity; and (d) viscosity.} \label{N2}
\end{figure*}

The real-fluid properties computed using the Enskog theory coupled with Virial EoS are first computed using different intermolecular potentials and Virial orders, aiming to reveal the fidelity needed in the intermolecular potentials and Virial orders to achieve accurate computation of real-fluid properties. Four cases are considered: (i) \textit{ab initio} intermolecular potential with $\rm{3^{rd}}$ order Virial EoS; (ii) \textit{ab initio} intermolecular potential with $\rm{2^{nd}}$ order Virial EoS; (iii) L-J intermolecular potential with $\rm{3^{rd}}$ order Virial EoS; and (iv) L-J intermolecular potential with $\rm{2^{nd}}$ order Virial EoS. Based on the results from our previous studies \cite{Wang2025} and the results in Fig. \ref{virial-all}, it can be expected that case (i) will demonstrate the best agreement with experiments. This is confirmed in Fig. \ref{N2} for both the thermodynamic and transport properties of $\rm{N_2}$. It can be seen from Fig. \ref{N2} that, with \textit{ab initio} intermolecular potential, obvious improvements can be achieved when increasing the Virial order from $\rm{2^{nd}}$ to $\rm{3^{rd}}$. The maximum error in the computed thermodynamic properties is below 2 \%, while the maximum error in the computed transport properties is below 11 \%.  It can be also seen from Fig. \ref{N2} that the performance of the Enskog theory with Virial EoS rapidly deteriorates when L-J potential is used. Regardless of the Virial order, the computed properties based on L-J potential deviate considerably from the experiments, both quantitatively and qualitatively. This is most obvious with thermal conductivity (i.e., Fig. \ref{N2}c), where the computed results based on L-J potential and $\rm{2^{nd}}$ order Virial EoS can be overestimated by approximately 100 \%. The same trends are also observed for other species, such as $\rm{H_2}$, $\rm{O_2}$ and $\rm{H_2O}$, as shown in Fig. S1, S2 and S3, respectively, in the Supplementary Material. These results emphasize that the application of Enskog theory with Virial EoS for computation of real-fluid transport properties, as well as the application of departure functions with Virial EoS for computation of real-fluid thermodynamic properties should be at least conducted with $\rm{3^{rd}}$ Virial order and shall not be derived based on L-J potentials, which corroborates the critical findings in our previous study \cite{Wang2025}.

\subsection{Ignition delay time}\label{section-3-2}

\begin{figure*}[h]
\centering
\includegraphics[width=\textwidth]{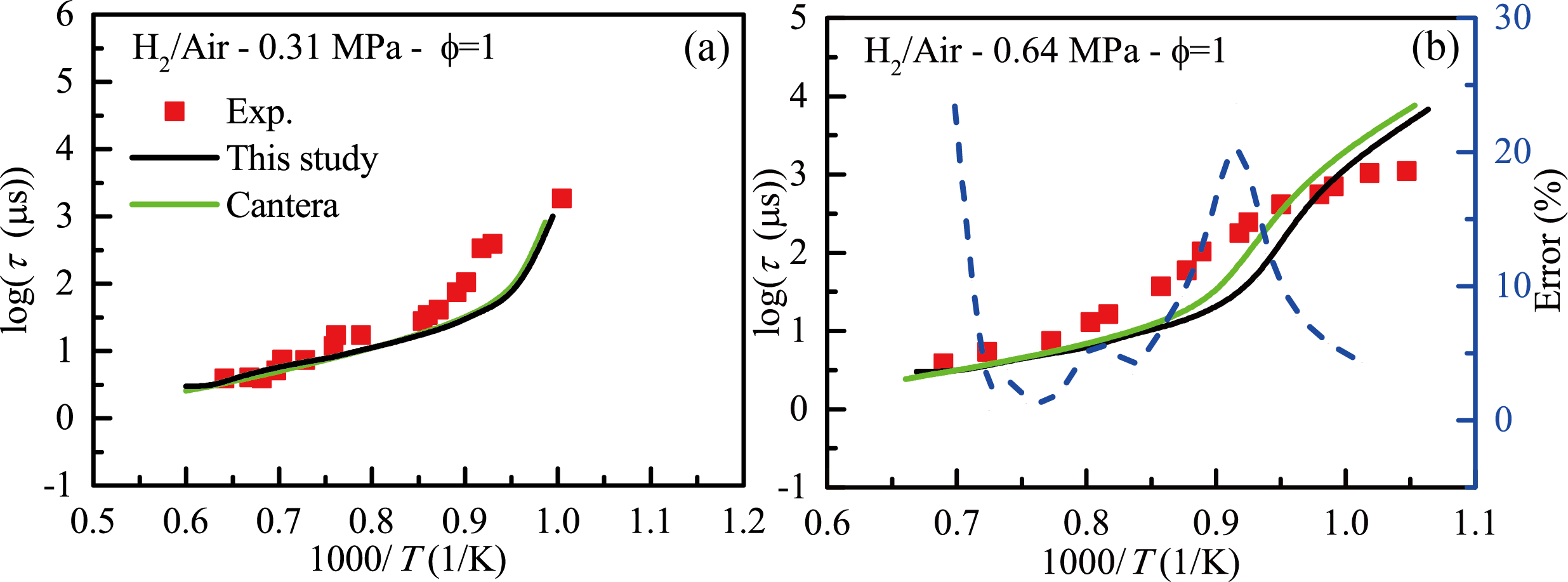}
\caption{Simulated IDTs of $\rm{H_2}$/Air mixture with and without considering real-fluid effects. (a) LFS at 0.31 MPa along with the experiments from \cite{Baranyshyn2024}; (b) IDT at 0.64MPa along with the experiments from \cite{Martynenko2004}. The ideal simulation results are marked in green; the real-fluid simulation results are marked in black, and the experiment results are marked with red squares.} \label{IDT}
\end{figure*}

As mentioned in Section \ref{section-2-4-1}, HiPrFlame can be used for 0-D to 3-D modeling with different settings for boundary conditions. To demonstrate the versatility of HiPrFlame, 0-D simulations are first conducted for autoignition in a homogeneous reactor \cite{Mao2023} with stoichiometric $\rm{H_2}$/air mixtures at 0.31MPa \cite{Baranyshyn2024} and 0.64MPa \cite{Martynenko2004}. Simulations with HiPrFlame are conducted with the $\rm{3^{rd}}$ order Virial EoS and \textit{ab initio} intermolecular potentials, while simulations without considering real-fluid behaviors are also conducted using Cantera \cite{goodwin2018cantera}. The results are summarized in Fig. \ref{IDT}, along with the experimental measurements from \cite{Baranyshyn2024, Martynenko2004}. It can be seen from Fig. \ref{IDT}a that, at 0.31 MPa, the simulation results from this study agree well with the ideal gas simulations using Cantera, indicating the minor real-fluid effects at this condition, which agree with our previous findings in \cite{Wang2025}. This also highlights the robustness of the developed framework for 0-D autoignition simulation. As pressure increases to 0.64 MPa (Fig. \ref{IDT}b), the real-fluid effects become pronounced at temperatures below 1100 K, where real-fluid effects increase autoignition reactivity, resulting in a reduction in ignition delay times.

\subsection{Laminar flame speed}
\begin{figure*}[!t]
\centering
\includegraphics[width=\textwidth]{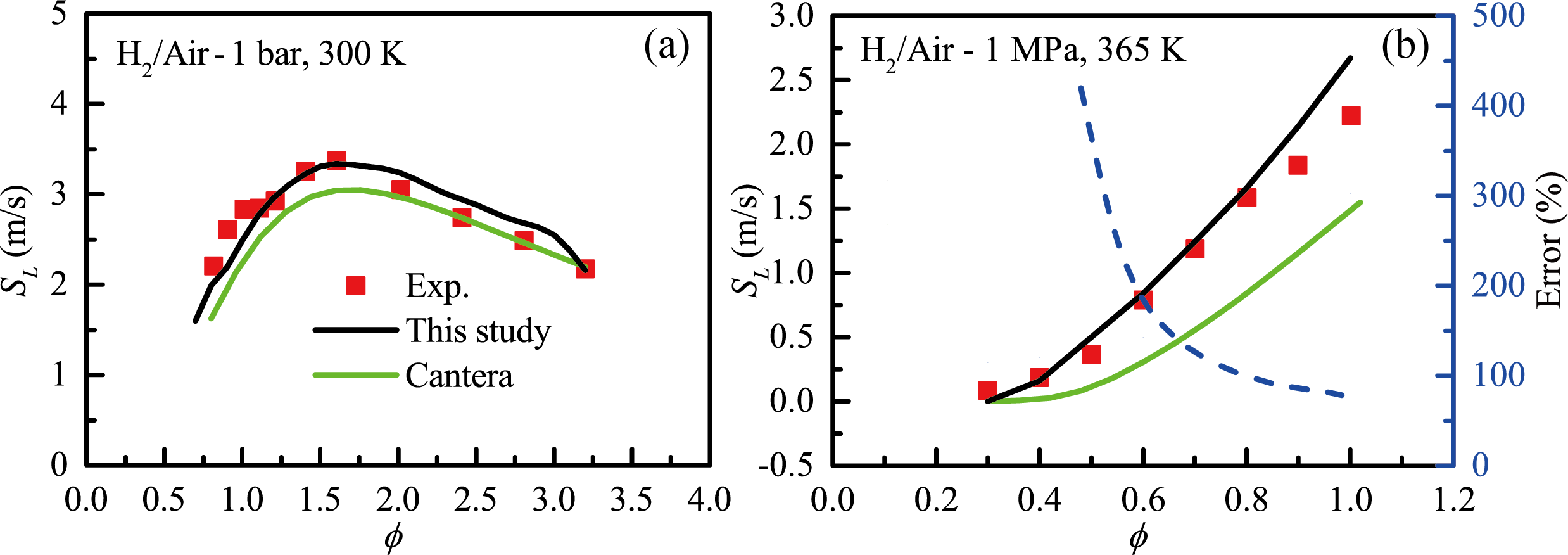}
\caption{Simulated LFSs of $\rm{H_2}$/Air mixtures with and without considering real-fluid effects. (a) LFS at 1 bar and 300 K along with the experiments from \cite{Ilbas2006}; (b) LFS at 1 MPa and 365 K along with the experiments from \cite{Bradley2007}. The ideal simulation results are marked in green, the real-fluid simulation results are marked in black and the experiment results are marked with red squares.} \label{flame_speed}
\end{figure*}

With the robustness of HiPrFlame demonstrated in 0-D autoignition simulations, HiPrFlame is further used for modeling 1-D laminar premixed flames of $\rm{H_2}$/air mixtures. The experimental conditions from two previous studies are adopted, where undiluted $\rm{H_2}$/air mixtures at 1 bar and 300 K \cite{Ilbas2006} and at 1.0 MPa and 365 K \cite{Bradley2007} were used. Similar to Section \ref{section-3-2}, simulations are also conducted using the $\rm{3^{rd}}$ order Virial EoS and \textit{ab initio} intermolecular potentials, with ideal gas simulation conducted using Cantera. The simulation results, along with the experimental measurements from \cite{Bradley2007,Ilbas2006}, are summarized in Fig. \ref{flame_speed}. As can be seen from Fig. \ref{flame_speed}, HiPrFlame replicates the ideal gas simulation results at 1 bar (Fig. \ref{flame_speed}a). Although slight differences are observed, both simulations agree well with the experimental measurements. However, at higher pressure (i.e., $p$ = 1.0 MPa, Fig. \ref{flame_speed}b) where real-fluid effects become more significant, the simulated LFSs from HiPrFlame display distinct increases as compared to the ideal gas simulations. The absolute increase in the simulated LFSs with HiPrFlame becomes greater at conditions closer to stoichiometry, reaching approximately 1 m/s at stoichiometry. It is also important to learn from Fig. \ref{flame_speed} that real-fluid effects are already impactful at conditions below the critical point (e.g., the pressure in Fig. \ref{flame_speed}b is 1.0 MPa, while the critical pressure of air is 3.77 MPa). The errors observed in Fig. \ref{flame_speed}b are too significant and much greater than those observed in 0-D simulations \cite{Wang2025,Wang2025a,Wang2025b,Wang2025d} (due most likely to the additional contributions from real-fluid transport), which shall be considered in modeling high-pressure flames. With real-fluid effects fully incorporated, the simulation results at the high-pressure condition achieve better agreement with the experiments than the ideal gas simulation, as can be seen from Fig. \ref{flame_speed}b.

\begin{figure}[h]
\centering
\includegraphics[scale=0.65]{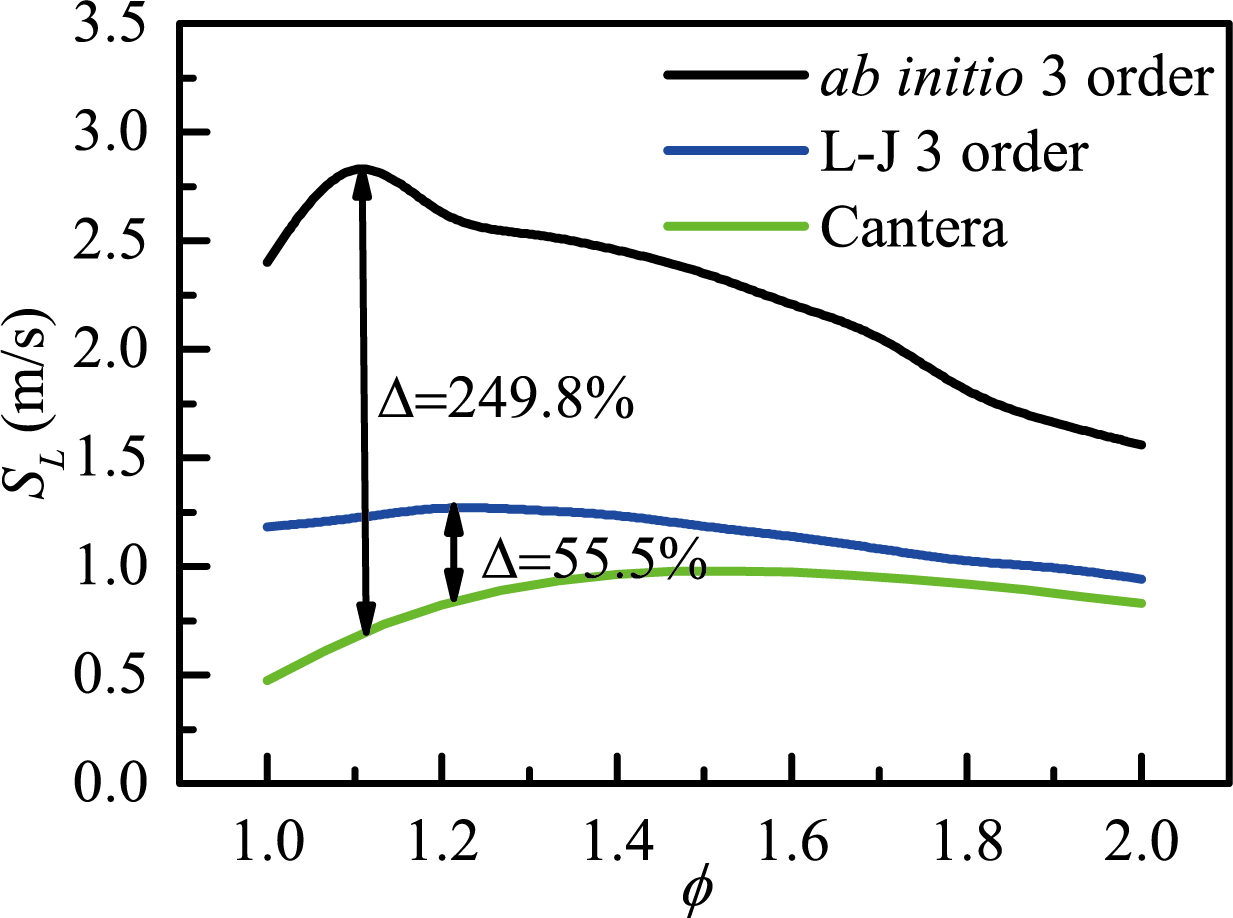}
\caption{LFS of $\rm{H_2}$/air mixtures at 300 K and 10 MPa, computed using the Enskog theory coupled with $\rm{3^{rd}}$ order Virial EoS based on different intermolecular potentials, along with the ideal gas simulations using Cantera.} \label{10MPa-lfs}
\end{figure}

The influences of intermolecular potential on the real-fluid modeling of high-pressure LFSs are further investigated. The two cases with $\rm{3^{rd}}$ order Virial EoS, same as those considered in Section \ref{section-3-1}, are investigated, and the simulated LFSs are illustrated in Fig. \ref{10MPa-lfs} for $\rm{H_2}$/air mixtures at 10 MPa and 300 K. First seen in Fig. \ref{10MPa-lfs} are the significant real-fluid effects at the studied conditions, which, again, increase the simulated LFSs, with greater impacts observed at equivalence ratios close to 1.1.  It can be further seen from Fig. \ref{10MPa-lfs} that with L-J potential, the real-fluid effects are considerably underestimated, e.g., by approximately 200 \% at equivalence ratio of 1.1. These results, again, highlight the superiority of HiPrFlame and the necessity of using \textit{ab initio} intermolecular potentials (rather than L-J potentials) for real-fluid modeling of high-pressure flames.

\section{Conclusions}
This study introduces HiPrFlame, an \textit{ab initio}-based modeling framework for high-pressure combustion that enables unprecedented accuracy in characterizing real-fluid effects. The following points summarize the analyses presented in this paper, which form the foundation of HiPrFlame:

1.	HiPrFlame achieves high-fidelity representation of real-fluid behavior by leveraging real-fluid partition function theory (via the third-order Virial equation of state) and non-ideal molecular interactions (via \textit{ab initio} intermolecular potentials). Using this approach, second and third Virial coefficients are computed from \textit{ab initio} potentials with advanced sampling techniques, outperforming the NIST database and demonstrating substantial superiority over empirical potentials in reproducing experimental Virial coefficient data.

2.	HiPrFlame employs real-fluid departure functions in conjunction with the third-order Virial EoS to characterize thermodynamic properties, achieving excellent agreement with experimental measurements and significantly surpassing the accuracy of cubic EoS models.

3.	For the first time, HiPrFlame enables the coupling of Enskog theory with an \textit{ab initio}-based third-order Virial EoS to characterize real-fluid transport properties in high-pressure combustion. This approach delivers markedly improved performance compared to existing empirical methods.

4.	HiPrFlame is architected in OpenFOAM with versatility for zero- to three-dimensional modeling, depending on the specified boundary conditions, where the correction of real-fluid properties is achieved via a front-end updating mechanism that couples ANN surrogate models trained on optimized machine learning frameworks.

This approach is demonstrated via a case study of high-pressure hydrogen combustion, including both homogeneous autoignition and one-dimensional laminar premixed flames. Results show that real-fluid effects enhance reactivity, significantly advancing ignition timing and increasing laminar flame speeds. Notably, with the inclusion of real-fluid transport, simulated laminar flame speeds can be enhanced by more than 400 \% at pressures even below the critical point (e.g., 1 MPa), a much greater effect than previously observed in homogeneous reactors. Both the real-fluid property predictions and combustion modeling results underscore the superiority of HiPrFlame and emphasize that, for accurate computation of real-fluid transport and thermodynamic properties, Enskog theory and departure functions, respectively, should be applied with \textit{ab initio} intermolecular potentials and at least third-order Virial EoS, and shall not be based on L-J potentials.

\section*{CrediT authorship contribution statement}

\textbf{Ting Zhang:} Methodology, Software, Investigation, Writing-original draft. \textbf{Tianzhou Jiang:} Investigation, Writing-review \& editing. \textbf{Mingrui Wang:}  Investigation, Writing-review \& editing. \textbf{Hongjie Zhang:} Investigation, Visualization. \textbf{Ruoyue Tang:} Investigation, Visualization, Writing-review \& editing. \textbf{Xinrui Ren:} Investigation, Visualization, Writing-review \& editing. \textbf{Song Cheng:} Conceptualization, Methodology, Investigation, Writing-review \& editing, Supervision, Project administration, Funding acquisition. 

\section*{Declaration of competing interest}

The authors declare that they have no known competing financial interests or personal relationships that could have appeared to influence the work reported in this paper.

\section*{Acknowledgments}

The work described in this paper is supported by the Research Grants Council of the Hong Kong Special Administrative Region, China under 25210423 for ECS project funded in 2023/24 Exercise, the National Natural Science Foundation of China under 52406158, and the Natural Science Foundation of Guangdong Province under 2024A1515011486. This work was supported by the High-Performance Computing resources provided by the University Research Facility in Big Data Analytics (UBDA) of the Hong Kong Polytechnic University.

\section*{Supplementary material}

Supplementary Material is submitted along with the manuscript.

\FloatBarrier

\bibliographystyle{cnf-num}
\bibliography{HiPrFlame}

\end{document}